\documentclass[sigconf]{acmart}

\usepackage{booktabs} 
\usepackage{boldline}

\usepackage{lipsum} 
\usepackage{balance}
\usepackage{paralist}
\usepackage{filecontents}
\usepackage{etoolbox}
\usepackage{mathrsfs}
\usepackage{bm}
\usepackage{setspace}
\usepackage{amssymb}
\usepackage{graphicx}
\usepackage{caption}
\usepackage{subcaption}
\usepackage{multirow}
\copyrightyear{2018} 
\acmYear{2018} 
\setcopyright{acmcopyright}
\acmConference[SIGIR '18]{The 41st International ACM SIGIR Conference on Research and Development in Information Retrieval}{July 8--12, 2018}{Ann Arbor, MI, USA}
\acmBooktitle{SIGIR '18: The 41st International ACM SIGIR Conference on Research and Development in Information Retrieval, July 8--12, 2018, Ann Arbor, MI, USA}
\acmPrice{15.00}
\acmDOI{10.1145/3209978.3209985}
\acmISBN{978-1-4503-5657-2/18/07}

\begin{document}

\title{Learning a Deep Listwise Context Model for Ranking Refinement}
%
%
%
%
%

\author{Qingyao Ai$^1$,Keping Bi$^1$, Jiafeng Guo$^2$,  W. Bruce Croft$^1$}
\affiliation{%
	\institution{$^1$College of Information and Computer Sciences, University of Massachusetts Amherst}
	\city{Amherst} 
	\state{MA} 
	\country{USA}
	\postcode{01003-9264}
}
\email{{aiqy, kbi, croft}@cs.umass.edu}
\affiliation{%
	\institution{$^2$CAS Key Lab of Network Data Science and Technology, Institute of Computing Technology, \\ Chinese Academy of Sciences}
	\city{Beijing} 
	\country{China} 
	\postcode{100190}
}
\email{guojiafeng@ict.ac.cn}

\begin{abstract}

Learning to rank has been intensively studied and widely applied in information retrieval. 
Typically, a global ranking function is learned from a set of labeled data, which can achieve good performance on average but may be suboptimal for individual queries by ignoring the fact that relevant documents for different queries may have different distributions in the feature space.
Inspired by the idea of pseudo relevance feedback where top ranked documents, which we refer as the \textit{local ranking context}, can provide important information about the query's characteristics, we propose to use the inherent feature distributions of the top results to learn a Deep Listwise Context Model that helps us fine tune the initial ranked list. 
Specifically, we employ a recurrent neural network to sequentially encode the top results using their feature vectors, learn a local context model and use it to re-rank the top results. 
There are three merits with our model: (1) Our model can capture the local ranking context based on the complex interactions between top results using a deep neural network; (2) Our model can be built upon existing learning-to-rank methods by directly using their extracted feature vectors; (3) Our model is trained with an attention-based loss function, which is more effective and efficient than many existing listwise methods.
Experimental results show that the proposed model can significantly improve the state-of-the-art learning to rank methods on benchmark retrieval corpora.

\end{abstract}

%
%


\maketitle

\section{Introduction}

Ranking is a core problem of information retrieval (IR).
Many IR applications such as ad-hoc retrieval, summarization and recommendations are ranking problems by nature~\cite{liu2009learning}.
Among all the ranking paradigms, learning to rank is the most widely used technology in modern search systems.
The idea of learning to rank is to represent each object with a manually designed feature vector and learn a ranking function with machine learning techniques. 
In document retrieval, for example, the ranking objects are query-document pairs and the vector representation of a query-document pair usually consists of multiple document or query features such as BM25 scores, click through rates, query quality scores etc. 
The ranking functions are typically learned globally on labeled query-document pairs from a separate training dataset~\cite{burges2005learning,joachims2006training,liu2009learning,quoc2007learning,cao2007learning}.

Such a global ranking function, however, may not be optimal for document retrieval as it ignores the differences between feature distributions for each query.
Depending on the query characteristics and user intents, relevant documents for different queries often have different distributions in feature space.
Considering two features such as word matching and freshness, relevant pages for a query like ``friends season 1 online watch" often have high scores on word matching but freshness is a lower priority; relevant documents for a query such as ``political news", on the other hand, should have high values of freshness but word matching scores are less important.
No matter how we design the feature vectors, these differences are inevitable and hard to solve with a global ranking function. 

A better paradigm for learning to rank is to learn a ranking model that can take into account the query-specific feature distributions.
Ideally, ranking functions would be constructed for each query separately~\cite{can2014incorporating, geng2008query}, but this would lead to unreasonable cost and low generalization ability because the number of possible queries is almost infinite and we do not know the feature distribution of an unseen query in advance.
As a compromise, a more practical method is to learn a local model for each query on the fly and use it to refine the ranking results.
For example, a well-studied framework is to represent each query with the top retrieved documents, namely the \textit{local ranking context}.
Previous studies~\cite{robertson1976relevance,salton1997improving,lavrenko2001relevance,zhai2004study} have shown that pseudo relevance models learned from the local ranking context can significantly improve the performance of many text-based retrieval models.

Given previous observations~\cite{lavrenko2001relevance,zhai2004study,qin2008global}, it seems intuitive to assume that the local context information from top ranked documents would benefit the performance of learning-to-rank systems.
Nonetheless, the utility of this information has not been fully studied.
One of the key challenges is how to develop a ranking model by using the feature representations of top results effectively and efficiently. 
On the one hand, there is no trivial solution to extract patterns from a group of feature vectors with hundreds of dimensions (which is common in modern search engines). 
Instead, most previous studies focus on constructing models using the text of documents alone~\cite{lavrenko2001relevance,zhai2001model} and ignore other ranking signals.
Those methods usually require an additional feature extraction (e.g. term extractions from top documents) and retrieval process in order to generate the final ranked list. 
On the other hand, re-ranking retrieved documents without considering their inherent structure could be risky.
Global information from the initial retrieval, namely the ranking positions of top results, is a strong indicator of document relevance and should be considered when we encode and fine-tune the ranked list for each query.  

To tackle these challenges, we propose a Deep Listwise Context Model (DLCM) that directly encodes the feature vectors of top retrieved documents to learn a local context embedding and use it to improve the learning-to-rank systems.
Specifically, we sequentially feed the original features of the top ranked results from a global learning-to-rank model into a recurrent neural network (RNN). 
The network state and the hidden outputs of the RNN are then used to re-rank the results from the initial retrieval.
There are several properties of our model that make it novel compared to previous studies. 
First, to the best of our knowledge, our model is the first model that directly incorporates the local ranking context from top results into a learning-to-rank framework. 
Second, our model uses the original feature representations and ranked lists from an existing system, which means that it can be directly deployed with most learning-to-rank models without additional term or feature extraction from the top retrieved documents. 
We adopt a re-ranking framework and require no additional retrieval process on document corpus after the initial run.
Last, we propose an attention-based listwise loss for the training of our model.
Models trained with our attention-based loss are more efficient and effective than those trained with traditional listwise loss functions such as ListMLE~\cite{xia2008listwise}. 

To demonstrate and understand the effectiveness of our model, we conducted empirical experiments on large-scale learning-to-rank corpora. 
Experimental results show that our model outperformed the state-of-the-art learning-to-rank algorithms significantly and consistently.
In addition, our analysis shows that our model was particularly good at finding the best document from a group of results, which potentially makes it useful for ranking scenarios where performance at high ranks is extremely important.

\section{Related Work}

There are two lines of previous studies related to our work: the research on learning-to-rank algorithms and the study of query-specific ranking.

\textbf{Learning to rank} refers to a group of techniques that attempts to solve ranking problems by using machine learning algorithms with the feature representations of query-document pairs.
The framework of learning to rank has been successfully applied in multiple areas such as question answering~\cite{yang2016beyond}, recommendation~\cite{duan2010empirical}, and document retrieval~\cite{joachims2002optimizing,liu2009learning}.

In document retrieval, the output of a learning-to-rank model is a score which indicates the relevance of a document for a query. 
Depending on how training losses are computed, learning-to-rank algorithms can be broadly categorized as pointwise, pairwise or listwise methods.
The pointwise methods treat the ranking problem as a classification or regression problem by taking one query-document pair a time and directly predicting its relevance score~\cite{friedman2001greedy}.
The pairwise methods transform the document ranking into a pairwise classification task by taking two documents a time and optimizing their relative positions in the final ranked list~\cite{joachims2006training,burges2005learning}.
The listwise methods further extend the above methods by taking multiple documents together and directly maximizing the ranking metrics~\cite{cao2007learning,taylor2008softrank,xia2008listwise,burges2010ranknet}.
For example, Taylor et al.~\cite{taylor2008softrank} trained models by optimizing the expected ranking metric computed with the expected rank of each document given a predefined Gaussian distribution. 

Recently, a couple of deep models have been proposed to extract features from raw data and predict the relevance of documents with neural networks~\cite{guo2016deep,dehghani2017neural,cheng2016wide,huang2013learning}.
For example, Guo et al.~\cite{guo2016deep} built matching histograms for each query-document pair and train a deep neural network to predict their relevance.
Despite the differences in loss functions and model structures, all these models try to learn a global ranking function that predicts the relevance score of a document purely based on its own feature representation.
They assume that the feature vectors of relevant query-document pairs are sampled from a global distribution and ignore the fact that documents for different queries may have different feature distributions.
As a contrast, we propose to use the local context from the top retrieved documents to model the query-specific feature distributions and conduct re-ranking accordingly.


\textbf{Query-specific ranking}.
The best way to do query-specific ranking is to build ranking schema for each query independently. 
Training models for each query separately, however, is not feasible in practice because we do not have labeled data for unseen queries.
As a compromise, previous studies learned multiple ranking models on training sets and rank documents for test queries by using the pre-constructed models for similar training queries~\cite{geng2008query,can2014incorporating}.
For example, 
Can et al.~\cite{can2014incorporating} constructed individual ranking models for each training query and aggregated the model scores according to the similarity between training queries and the test query. 
In contrast to these studies, the core of our work is not to find similar queries in the training set but to directly model and use the local ranking context of each test query on the fly.

Another research direction focuses on extracting features from the top retrieved documents to improve the initial ranking.
A well-known technique is pseudo relevance feedback~\cite{lavrenko2001relevance,zhai2001model}.
For example, Lavrenko and Croft~\cite{lavrenko2001relevance} treated each document in the top results as a unigram distribution and sum over the joint probability of observing a word together with the query to form a relevance model for query expansion.
Zhai and Lafferty~\cite{zhai2001model} extracted a topic model from the words in feedback documents and interpolated it into the original query model.
Compared to text-based retrieval models, there has not been much work on using the top retrieved documents for learning-to-rank algorithms.
To the best of our knowledge, the only studies in this area are the CRF-based ranking model~\cite{qin2008global} and the score regularization technique~\cite{diaz2007regularizing}.
Different from our work, both of them focus on utilizing the document similarity features computed with word distributions but not the modeling of local ranking context.
They are expensive and limited because they require accessing the raw text of documents after the initial retrieval for feature extraction and ignore document relationships based on ranking features other than term vectors.
\section{Learning to Rank with The Local Ranking Context}\label{sec:LTR}

In this section, we formalize the problem of how to adapt the learning-to-rank frameworks with the local ranking context from top retrieved documents. 
Given a specific query $q$, a vector $\bm{x}_{(q,d)}$ can be extracted and used as the feature representation for a document $d$.
Traditional learning-to-rank algorithms assume that there exists an optimal global ranking function $f$ which takes $\bm{x}_{(q,d)}$ as its input and outputs a ranking score for the document.
The way to find this optimal $f$ is to minimize a loss function $\mathcal{L}$ defined as
\begin{equation}
\mathcal{L} = \sum_{q \in Q}\ell\big(\big\{y_{(q,d)},f(\bm{x}_{(q,d)}) \big| d \in D \big\}\big)
\label{equ:LTR}
\end{equation}
where $Q$ is the set of all possible queries, $D$ is the set of candidate documents, $\ell$ is the local loss computed with the document score $f(\bm{x}_{(q,d)})$ and corresponding relevance judgment $y_{(q,d)}$.
Now, suppose that we can capture the local ranking context of $q$ with a local context model $I(R_q, X_q)$ where $R_q = \{d \text{  sorted by } f(\bm{x}_{(q,d)})\}$ and $X_q = \{\bm{x}_{(q,d)}|d\in R_q\}$, then the loss of learning to rank with local context can be formulated as:
\begin{equation}
\mathcal{L} = \sum_{q \in Q}\ell\bigg(\big\{y_{(q,d)},\phi\big(\bm{x}_{(q,d)},I(R_q, X_q)\big)\big| d \in D \big\}\bigg)
\label{equ:LTR_li}
\end{equation}
where $\phi$ is a scoring function that ranks documents based on both their features and the local context model $I(R_q, X_q)$.
The goal is to find the optimal $I$ and $\phi$ that minimize the loss function $\mathcal{L}$.


To effectively utilize the local ranking context, the design of the listwise context model $I$ should satisfy two requirements. 
First, it should be able to process scalar features directly. 
Most of the learning-to-rank systems convert ranking signals, whether discrete or continuous, to a vector of scalar numbers.
If the listwise context model $I$ cannot deal with these scalar numbers directly, we need to extract the raw data from documents and manually develop heuristics to model the local ranking context, which is difficult and inefficient.
Second, it should consider the position effect of top retrieved documents. 
The value of documents in the top results is not the same and their positions ranked by the global ranking function are strong indicators of their relevance.
Without explicitly modeling the position effect, we would lose the global ranking information and harm the generalization ability of the whole system. 


\section{Deep Listwise Context Model}\label{sec:model}

\begin{figure*}
	\centering
	\includegraphics[width=5.5in]{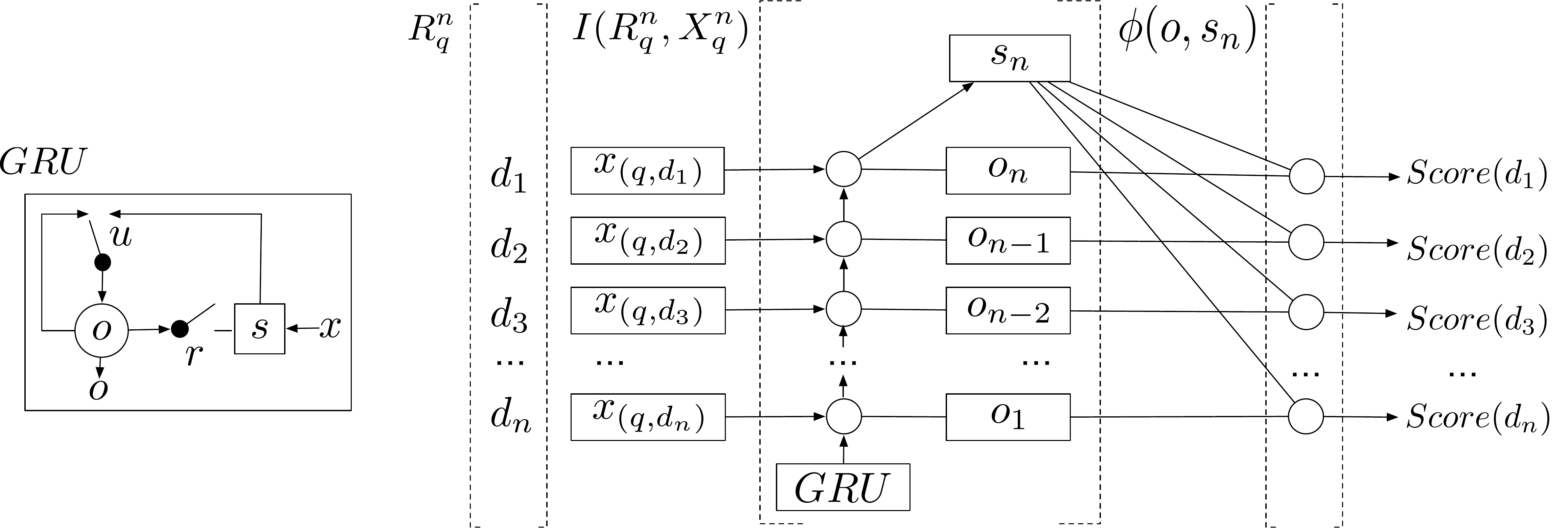}
	\vspace{-5pt}
	\caption{The overall structure of the Deep Listwise Context Model (DLCM). $R_q^n$ is a ranked list provided by a global ranking function $f$ for query $q$; $\bm{x}_{(q,d_i)}$ is the feature vector for document $d_i$; $\bm{s}_n$ and $\bm{o}_i$ is the final network state and hidden outputs of the RNN with GRU in $I(R_q^n,X_q^n)$; and $Score(d_i)$ is the final ranking score of $d_i$ computed with $\phi(\bm{o}_{n+1-i}, \bm{s}_n)$}
	\vspace{-5pt}
	\label{fig:our_model}
\end{figure*}

In this paper, we propose a deep neural model to incorporate the local ranking context into the learning-to-rank framework.
The overall idea of our model is to encode the top retrieved documents of each query with a recurrent neural network and refine the ranked list based on the encoded local context model.
We refer our model as the Deep Listwise Context Model (DLCM).

The pipeline of document ranking with DLCM includes three steps.
The first step is an initial retrieval with a standard learning-to-rank algorithm. 
In this step, each query-document pair $(q,d)$ is converted into a feature vector $\bm{x}_{(q,d)}$ and a ranked list $R_q^n$ with size $n$ is generated for query $q$ based on a global ranking function $f$.
The second step is an encoding process that uses a recurrent neural network (RNN) with gated recurrent unit (GRU) to encode the feature vectors $X_q^n$ of top retrieved documents.
The RNN takes documents one by one from the lowest position to the highest, and produces a latent vector $\bm{s}_n$ to represent the encoded local context model $I(R_q^n, X_q^n)$.
The third step is a re-ranking process where the top documents are re-ranked with a local ranking function $\phi$ based on both $\bm{s}_n$ and the hidden outputs $\bm{o}$ of the RNN.    
The overall structure of DLCM is shown in Figure~\ref{fig:our_model}.

\subsection{Input Document Representations}\label{sec:input}

As discussed in Section~\ref{sec:LTR}, most learning-to-rank algorithms use a feature vector to represent each query-document pair. 
In our proposed framework, the DLCM uses the same feature vectors as those used in previous learning-to-rank challenges \cite{DBLP:journals/corr/QinL13,chapelle2011yahoo}, which include both document and query related features.
We do not incorporate any additional features in the model inputs.

Directly feeding the original feature vectors into our model, however, may not be the best method to use the full strength of the neural network.
On one hand, the dimensionality of the original features may be limited and using low-dimensional representations would restrict the expressive ability of neural encoders.
On the other hand, high-level feature abstractions could be beneficial for the robustness of neural models especially when the original input features are noisy.
Inspired by the Wide\&Deep neural network \cite{cheng2016wide}, we apply a two-step method to obtain high-dimensional input representations for the DLCMs. 
We first use a two-layer feed-forward network to learn an abstraction of the original features:
\begin{equation}
\begin{split}
\bm{z_i^{(0)}} &= \bm{x}_{(q,d_i)} \\
\bm{z_i^{(l)}} &= elu(\bm{W}_z^{(l-1)} \cdot \bm{z}_i^{(l-1)} + \bm{b}_z^{(l-1)}), l = 1,2
\end{split}
\end{equation} 
where $\bm{W}_z^{(l)}$ and $\bm{b}_z^{(l)}$ are the weight matrix and bias in the $l$th layer and $elu$ is a non-linear activation function that equals to $x$ when $x \geq 0$ and $e^x-1$ otherwise.
We then concatenate $\bm{z}_i^{(2)}$ with the original feature vector $\bm{x}_{(q,d_i)}$ to form a new input vector $\bm{x}'_{(q,d_i)}$.
Let $\alpha$ and $\beta$ be the dimension of $\bm{x}_{(q,d_i)}$ and $\bm{z}_i^{(2)}$.
Because $\bm{x}'_{(q,d_i)}$ could be reduced to $\bm{x}_{(q,d_i)}$ when $\beta$ is equal to zero, we do not differentiate them in further discussions. 
  

\subsection{Encoding the Listwise Local Context}

Given the top $n$ results retrieved by a global ranking function $f$ and their corresponding feature vectors $X_q^n =\{\bm{x}_{(q,d_i)} | d_i \in R_q^n\}$, the local context model $I$ in DLCM is implemented with a recurrent neural network (RNN).
The RNN is a type of deep network widely used for the modeling of sequential data \cite{cho2014learning,sutskever2014sequence,vinyals2015pointer}.
A standard RNN consists of an input sequence, an output sequence and a state vector.
As we feed the input instances one by one (each instance is represented with a feature vector), the RNN updates its state vector according to current input and generates a new output vector in each step.
The final state vector can be viewed as an encoding of all the information that has been fed into the network.

In our DLCM, we use a RNN with gated recurrent unit (GRU).
The GRU network is a technique proposed by Cho et al.~\cite{cho2014properties} which aims to solve the problem of gradient vanishing in RNN.
Its basic idea is to control the update of network states with an update gate and a reset gate.
Formally, let $\bm{x}_t \in \mathbb{R}^\alpha$ be the input vector in the $t$ step and $\alpha$ is the dimensionality of $\bm{x}_t$. 
The output vector (also the activation vector in GRU) $\bm{o}_t \in \mathbb{R}^\alpha$ and the network state $\bm{s}_t\in \mathbb{R}^\alpha$ is computed as:
\begin{equation}
\begin{split}
\bm{o}_t &= (1 - \bm{u}_t)\odot\bm{o}_{t-1} + \bm{u}_t\odot\bm{s}_t \\
\bm{u}_t &= \sigma(\bm{W}_u^x \cdot \bm{x}_t + \bm{W}_u^s \cdot \bm{o}_{t-1}) \\
\bm{s}_t &= \tanh\big(\bm{W}^x \cdot \bm{x}_t + \bm{W}^s \cdot (\bm{r}_t\odot\bm{o}_{t-1})\big)\\
\bm{r}_t &= \sigma(\bm{W}_r^x\cdot \bm{x}_t + \bm{W}_r^s\cdot \bm{o}_{t-1})
\label{equ:state_t}
\end{split}
\end{equation}
where $\odot$ is the element-wise product, $\sigma(x) = \frac{1}{1 + e^{-x}}$ is a sigmoid function, $\bm{u}_t \in \mathbb{R}^\alpha$ is the update gate and $\bm{r}_t\in \mathbb{R}^\alpha$ is the reset gate.
All weight matrices $\bm{W}^x, \bm{W}^s, \bm{W}_u^x, \bm{W}_u^s, \bm{W}_r^x, \bm{W}_r^s \in \mathbb{R}^{\alpha \times \alpha}$ are learned in the training process. 
The encoded local context model $I(R_q^n,X_q^n)$ is the final network state $\bm{s}_n$.

The RNN with GRU in the DLCM naturally satisfies the two requirements of a local context model as discussed in Section~\ref{sec:LTR}.
The inputs of RNN are a sequence of vectors, which could be the scalar features from learning-to-rank systems. 
Because the RNN automatically learns to combine the current input with previous inputs encoded in the network state, we do not need to manually define heuristics to model the local ranking context. 
Also, the structure of the RNN enables it to capture the position effect in the encoding process.
When we feed the network with input data one by one, the current input tends to have more influence on the current network state than previous inputs. 
Since we input the sorted top results from the lowest position to the highest, documents in the high positions will have more impact on the final network state.

As an alternative to the uni-directional RNN shown in Figure~\ref{fig:our_model}, we also tested the bi-directional RNN~\cite{schuster1997bidirectional}.
Although it is considered to be more advanced in NLP tasks~\cite{bahdanau2014neural}, we observed no improvement in our retrieval experiments when replacing the uni-directional RNN with the bi-directional one. 
This indicates that the information encoded from the reversed direction is not useful.
In fact, if we only used the uni-directional RNN on the reversed direction, the performance of the DLCM would be significantly worse.

\subsection{Re-ranking with the Local Context}

The final step of ranking in our DLCM is to produce a new ranked list by sorting documents with a local ranking function $\phi$.
When predicting a ranking score, the function $\phi$ considers both the hidden outputs of RNN and the encoded latent representation of the local ranking context.
Let $\bm{o}_{n+1-i}$ be the output representation of document $d_i \in R_q^n$, we define the local ranking function $\phi$ as
\begin{equation}
\phi(\bm{o}_{n+1-i}, \bm{s}_n) = \bm{V}_{\phi} \cdot \big(\bm{o}_{n+1-i}\cdot \tanh(\bm{W}_{\phi} \cdot \bm{s}_n + \bm{b}_{\phi})\big)
\label{equ:phi}
\end{equation}
where $\bm{W}_{\phi} \in \mathbb{R}^{\alpha \times k \times \alpha}$, $\bm{b}_{\phi} \in \mathbb{R}^{\alpha \times k}$, and $\bm{V}_{\phi} \in \mathbb{R}^{k}$ and $k$ is a hyper-parameter that controls the number of hidden units. 

Our definition of the local ranking function is similar to the attention function widely used in RNN research~\cite{cho2014learning,vinyals2015pointer,merity2016pointer}.
In many machine learning applications (i.e. machine translation), a RNN decoder needs to pay attention to different part of the input data in different steps. 
For example, we need to focus on different parts of a sentence when we generate a translation word by word.
Attention functions are commonly used to compute an attention distribution over the input data and generate an attention vector to guide the decoding process of a RNN.
In the DLCM, we directly use the output values of $\phi$ to rank input documents. 

We tried other settings like replacing $\bm{o}$ with $\bm{x}$ and implementing $\phi$ with a three-layer feed forward network or a neural tensor network \cite{socher2013reasoning}. 
However, their performance were either worse or not significantly better than our method in the experiments, so we only report the results of $\phi$ defined as Equation~\ref{equ:phi} in this paper.

\subsection{Loss Function}

To train our DLCM, we implemented two existing listwise loss functions (\textit{ListMLE}~\cite{xia2008listwise} and \textit{SoftRank}~\cite{taylor2008softrank}) and also proposed a new listwise loss function called \textit{Attention Rank}.

\textit{\textbf{ListMLE}} is a listwise loss function that formulates learning to rank as a problem of minimizing the likelihood loss~\cite{xia2008listwise}.
It treats ranking as a sequential selection process and defines the probability to select document $d_i$ from documents $\pi_m^n=\{d_j | j\in[m,n]\}$ as
\begin{equation}
	P(d_i|\pi_m^n) =  \frac{e^{S_i}}{\sum_{j=m}^ne^{S_j}}
\end{equation}
where $S_i$ and $S_j$ are the ranking scores of $d_i$ and $d_j$.
If we start the selection from the top of a ranked list $R^n_q$ and remove the selected document from the candidate set after each step, we have the probability of observing $R^n_q$ given the ranking scores $\bm{S}$ as
\begin{equation}
P(R^n_q|\bm{S}) = \prod_{i=1}^{n} P(d_i|\pi_i^n) =  \prod_{i=1}^{n}\frac{e^{S_i}}{\sum_{j=i}^ne^{S_j}}
\end{equation}
Let $\mathcal{R}^{*}_q$ be the best possible ranked list for query $q$, then the ListMLE loss is defined as the minus of the log likelihood of $\mathcal{R}^{*}_q$ given $\bm{S}$.

\textit{\textbf{SoftRank}},
firstly proposed by Taylor et al.~\cite{taylor2008softrank}, is a listwise loss function that directly optimizes the ranking metrics of information retrieval such as NDCG.
Let $S_i$ and $S_j$ be the ranking scores of document $d_i$ and $d_j$ for query $q$. 
The SoftRank function assumes that the ``real" score $S_i'$ of document $d_i$ is drawn from a Gaussian distribution defined as $\mathcal{N}(S_i, \sigma_s^2)$ where $\sigma_s$ is a shared smoothing variance.
Given this assumption, the probability that $d_i$ is ranked higher than $d_j$ can be computed as:
\begin{equation}
\pi_{ij} \equiv \Pr(S_i' - S_j' > 0) = \int_{0}^{\infty} \mathcal{N}(S|S_i-S_j, 2\sigma_s^2)dS
\end{equation}
Let $p_j^{(1)}(r)$ be the initial rank distribution for $d_j$ when $d_j$ is the only document in the ranked list, then $p_j^{(i)}(r)$ after adding the $i$th document is computed as:
\begin{equation}
p_j^{(i)}(r) = p_j^{(i-1)}(r-1)\pi_{ij} + p_j^{(i-1)}(r)(1-\pi_{ij})
\end{equation}
With the final rank distribution $p_j^{(n)}(r)$ and the label of all $n$ documents, we can compute the expected relevance value on each rank and define a loss function as the minus of an expected metric score.

In this paper, we use NDCG as the objective metric for SoftRank. 
The only hyper-parameter in SoftRank is the shared smoothing variance $\sigma_s$.
We tried 0.1, 1.0 for $\sigma_s$ and observed no significant difference in respect of the retrieval performance. 
Therefore, we only report the results with $\sigma_s$ equal to 0.1.


\textit{\textbf{Attention Rank}}.
Inspired by previous work on attention-based neural networks, we propose an Attention Rank loss function that formulates the evaluation of a ranked list as a process of attention allocation.
Assuming that the information contained in documents is mutually exclusive, the total information gain of a ranked list is the accumulation of each document's gain.
If we further assume that the relevance judgment scores of a document directly reflect its information gain, the best strategy to maximize the total information gain in a limited time is to allocate more attention to the best results, less attention to the fair results and no attention to the irrelevant results.
The idea of Attention Rank is to compute an attention distribution with the ranking scores of our models and compare it with the attention strategy computed with the relevance judgments.
Let the relevance label $y_{(q,d_i)}$ represent the information gain of document $d_i$ for query $q$.
The best attention allocation strategy on a ranked list $R_q^n$ is defined as
\begin{equation}
a_i^y =  \frac{\psi(y_{(q,d_i)})}{\sum_{d_k \in R_q^n}\psi(y_{(q,d_k)})}
\end{equation} 
where $\psi(x)$ is a rectified exponential function that equals to $e^x$ when $x>0$ and equals to 0 otherwise.
Similarly, we compute the attention distribution of our model $a_i^S$ with the ranking score $S_i$ and use the cross entropy between our attention strategy and the best attention strategy as the loss of $R_q^n$:
\begin{equation}
\ell(R_q^n) = -\sum_{d_i \in R_q^n}\big(a^y_i\log(a^S_i) + (1-a^y_i)\log(1-a^S_i)\big)
\end{equation}

Attention Rank does not directly predict the relevance labels of documents but focuses on the relative importance of each result in the ranked list.
For example, a fair document in a list of irrelevant results could receive more attention than an excellent document in a list of perfect results.
Because it computes ranking loss based on ranked lists, Attention Rank is a listwise function rather than a pointwise function. 
The main advantages of Attention Rank are its simplicity and efficiency.
By using the rectified exponential function $\psi(x)$, we explicitly allocate more effort to optimize high-relevance results in the training process.
The training of the DLCM with Attention Rank was 2 and 20 times faster than the DLCM with ListMLE and SoftRank in our experiments. 
Also, it can be directly applied in the unbiased learning to rank framework~\cite{ai2018unbiased}.

\section{Experimental Setup}


In our experiments, we used three benchmark datasets, Microsoft 30k, Microsoft 10k~\cite{DBLP:journals/corr/QinL13}\footnote{ \url{https://www.microsoft.com/en-us/research/project/mslr/}} and Yahoo! Webscope v2.0 set 1\footnote{\url{http://webscope.sandbox.yahoo.com}}.
As far as we know, these are the largest public learning-to-rank datasets from commercial English search engines.
The statistics of corpora are listed in Table~\ref{tab:dataset}.
Due to privacy concerns, these datasets do not disclose any text information and only provide feature vectors for each query-document pair.
The Microsoft datasets are partitioned into five folds and define cross validation by using three folds for training, one fold for validation and one fold for testing.
The Yahoo! set 1 splits the queries arbitrarily and uses 19,944 for training, 2,994 for validation and 6,983 for testing. 

\begin{table}
	\centering
	\small
	\caption{The characteristics of learning-to-rank datasets used in our experiments: number of queries, documents, relevance levels, features and year of release.}
	\begin{tabular}{ l | l l l l l    } 
		\hline
		& Queries & Doc. & Rel. & Feat. & Year\\ \hline
		Micrsoft 30K & 31,531 & 3,771k & 5 & 136 & 2010 \\
		Micrsoft 10K & 10,000 & 1,200k & 5 & 136 & 2010 \\
		Yahoo! set 1 & 29,921 & 710k & 5 & 700 & 2010 \\
		\hline
	\end{tabular}
	\label{tab:dataset}
\end{table}

\textit{\textbf{Baselines}}.
We used two types of global learning-to-rank models as our baselines: SVMrank and LambdaMART.
SVMrank~\cite{joachims2006training} is a well-known ranking model trained with pairwise losses while LambdaMART~\cite{burges2010ranknet} is the state-of-the-art learning-to-rank algorithm trained with listwise losses.  
In this paper, we used the implementation of SVMrank\footnote{\url{https://www.cs.cornell.edu/people/tj/svm_light/svm_rank.html}} from Joachims~\cite{joachims2006training} and the implementations of LambdaMART from RankLib\footnote{\url{https://sourceforge.net/p/lemur/wiki/RankLib/}}.

We used global ranking algorithms to do the initial retrieval and showed the results of the DLCMs with three loss functions, the ListMLE, the SoftRank and the Attention Rank function (AttRank).
To demonstrate the effectiveness of the DLCM as a re-ranking model, we include three baselines that use global ranking models to re-rank the initial results.
They are the original ListMLE model~\cite{xia2008listwise}, the original SoftRank model~\cite{taylor2008softrank} and the model trained with our proposed AttRank loss (AttRank).
Because they all implement the ranking function with feed-forward Deep Neural Network (DNN), we refer to these three baselines as DNN with ListMLE, SoftRank and AttRank respectively.

As an initial attempt to use the local ranking context for learning to rank, we are interested to see how the DLCM performs compared to simple models that directly uses all the features from a ranked list.
Therefore, we incorporate a new baseline that concatenates the features of all documents as the inputs to a feed-forward neural network and predicts their ranking scores together. 
For example, suppose that the input ranked list $R_q^n$ has 40 documents ($n=40$) and each document has 700 features ($\alpha=700$), then the model takes a 28,000-dimension vector as its inputs and outputs a 40-dimension score vector. 
To differentiate it from models that score one document a time, we call this new baseline as the \textit{Listwise Input Deep Neural Network} (LIDNN).
Similar to the DLCM, we implemented LIDNN with ListMLE, SoftRank and AttRank respectively.


Due to the limitations of the datasets, we cannot access the text data of queries and documents, which makes it impossible to construct relevance models, extract document relationship features or compute query similarities.
The goal of our work is to improve learning-to-rank systems with the local ranking context but not to extract new features or design new models for query expansion. 
Therefore, we did not include models that use the raw text of queries or documents as baselines in our experiments~\cite{lavrenko2001relevance,diaz2007regularizing,qin2008global,geng2008query,can2014incorporating}.

\textit{\textbf{Evaluation}}.
Our datasets have five-level relevance judgments, from 0 (irrelevant) to 4 (perfectly relevant), so we use two types of multi-label ranking metrics.
The first one is the Normalized Discounted Cumulative Gain (NDCG)~\cite{jarvelin2002cumulated},
and the second one is Expected Reciprocal Rank (ERR)~\cite{chapelle2009expected}.
For both NDCG and ERR, we reported results at rank 1,3,5 and 10 to show the performance of our models on different positions.
Statistical differences are computed with the Fisher randomization test~\cite{smucker2007comparison} ($p \leq 0.01$).

\textit{\textbf{Model training}}.
The training of the DLCMs and the baselines includes two parts: the training of global ranking functions for the initial retrieval, and the training of ranking models for re-ranking.
We tuned the global ranking model on the validation set based on NDCG@10 and select the best one as our initial ranking function (which is also the baseline reported in this paper). 
For SVMrank, we tuned parameter $c$ from 20 to 200; for LambdaMART, we tuned tree number from 100 to 1000. 
For the re-ranking baselines and the LIDNNs, we tried both two-layer and three-layer feedforward neural networks with hidden layer units from 64 to 1024. 
We only reported the best results for each baseline.

For the training of RNN and local ranking functions, we used stochastic gradient descent with batch size 256. 
The initial learning rate is 1.0, and it decays by 0.8 each time when the training loss increases.
For each iteration, we randomly sampled a batch of queries to feed the model and clip the global gradient norm with 5 before update.
We trained our models on one Nvidia Titan X GPU with 12 GB memory. 
The training of the DLCMs with AttRank and ListMLE (10,000 iterations) takes about 2 to 4 hours while the training of the DLCMs with SoftRank usually takes 2 to 3 days. 
The re-ranking process usually takes about 2 to 3 ms for each test query.

There are three hyper-parameters for our DLCMs: the size of input ranked list $n$, the dimensions of input abstraction $\beta$ (Section~\ref{sec:input}), and the hidden unit number $k$ (Equation~\ref{equ:phi}). 
We tuned $n$ from 10 to 60, $\beta$ from 0 to 200, and $k$ from 1 to 15. 
The source code can be found in the link below\footnote{\url{https://github.com/QingyaoAi/Deep-Listwise-Context-Model-for-Ranking-Refinement}}.

\begin{table*}[t]
	\centering
	\small
	\caption{Comparison of baselines and the DLCMs on Micrsoft 30K. $*$, $+$ and $\ddagger$ denotes significant improvements over the global ranking algorithm and the best corresponding re-ranking baseline (DNN) and LIDNN. 
	}
	\scalebox{1.0}{
		\begin{tabular}{ c | c | c || l | l | l | l | l | l | l | l    } 
			\hline
			\multicolumn{3}{c||}{ } & \multicolumn{8}{c}{Microsoft Letor Dataset 30K}\\ \hline 
			Initial List & Model & Loss Function & nDCG@1 & ERR@1 & nDCG@3 & ERR@3 & nDCG@5 & ERR@5 & nDCG@10 & ERR@10  \\\hline \hline
			\multicolumn{3}{c||}{SVMrank} & 0.301 & 0.124 & 0.318 & 0.197 & 0.335 & 0.223 & 0.365 & 0.246 \\ \hline \hline
			\multirow{9}{*}{SVMrank}&\multirow{3}{*}{DNN} & ListMLE & 0.337$^{*\ddagger}$ & 0.149$^{*\ddagger}$ & 0.345$^{*\ddagger}$ & 0.224$^{*\ddagger}$ & 0.356$^{*\ddagger}$ & 0.249$^{*\ddagger}$ & 0.382$^{*\ddagger}$ & 0.271$^{*\ddagger}$ \\ \cline{3-11}
			& & SoftRank & 0.388$^{*\ddagger}$ & 0.208$^{*\ddagger}$ & 0.376$^{*\ddagger}$ & 0.279$^{*\ddagger}$ & 0.379$^{*\ddagger}$ & 0.300$^{*\ddagger}$ & 0.395$^{*\ddagger}$ & 0.318$^{*\ddagger}$ \\ \cline{3-11}
			& & AttRank & 0.395$^{*\ddagger}$ & 0.198$^{*\ddagger}$ & 0.392$^{*\ddagger}$ & 0.274$^{*\ddagger}$ & 0.396$^{*\ddagger}$ & 0.297$^{*\ddagger}$ & 0.415$^{*\ddagger}$ & 0.316$^{*\ddagger}$ \\ \cline{2-11} 
			& \multirow{3}{*}{LIDNN} & ListMLE & 0.291 & 0.122 & 0.312 & 0.196 & 0.331 & 0.222 & 0.362 & 0.245 \\ \cline{3-11}
			& & SoftRank & 0.315$^{*}$ & 0.141$^{*}$ & 0.326$^{*}$ & 0.213$^{*}$ & 0.341$^{*}$ & 0.238$^{*}$ & 0.367$^{*}$ & 0.260$^{*}$ \\ \cline{3-11}
			& & AttRank & 0.306$^{*}$ & 0.135$^{*}$ & 0.318 & 0.206$^{*}$ & 0.331 & 0.231$^{*}$ & 0.361 & 0.253$^{*}$ \\ \cline{2-11} 
			& \multirow{3}{*}{\textbf{DLCM}} & ListMLE &   0.339$^{*\ddagger}$ & 0.149$^{*\ddagger}$ & 0.346$^{*\ddagger}$ & 0.223$^{*\ddagger}$ & 0.357$^{*\ddagger}$ & 0.248$^{*\ddagger}$ & 0.381$^{*\ddagger}$ & 0.269$^{*\ddagger}$ \\ \cline{3-11}
			& & \textbf{SoftRank}  & \textbf{0.424}$^{*+\ddagger}$ & \textbf{0.224}$^{*+\ddagger}$ & \textbf{ 0.404}$^{*+\ddagger}$ & \textbf{0.294}$^{*+\ddagger}$ & \textbf{0.408}$^{*+\ddagger}$ & \textbf{0.316}$^{*+\ddagger}$ & \textbf{0.423}$^{*+\ddagger}$ & \textbf{0.334}$^{*+\ddagger}$ \\ \cline{3-11}
			& & AttRank &   0.407$^{*+\ddagger}$ & 0.206$^{*\ddagger}$ & 0.399$^{*+\ddagger}$ & 0.281$^{*+\ddagger}$ & 0.404$^{*+\ddagger}$ & 0.303$^{*+\ddagger}$ & 0.422$^{*+\ddagger}$ & 0.322$^{*+\ddagger}$ \\ \hline \hline
			\multicolumn{3}{c||}{LambdaMART} & 0.457$^{+}$ & 0.235$^{+}$ & 0.442$^{+}$ & 0.314$^{+}$ & 0.445$^{+}$ & 0.336$^{+}$ & 0.464$^{+}$ & 0.355$^{+}$ \\ \hline \hline 
			\multirow{9}{*}{LambdaMART} & \multirow{3}{*}{DNN} & ListMLE &0.372 & 0.174 & 0.378 & 0.254 & 0.386 & 0.278 & 0.409 & 0.299 \\ \cline{3-11}
			& & SoftRank & 0.384 & 0.209 & 0.373 & 0.281 & 0.378 & 0.302 & 0.397 & 0.321 \\ \cline{3-11}
			& & AttRank & 0.388 & 0.199 & 0.386 & 0.274 & 0.393 & 0.297 & 0.416 & 0.317 \\ \cline{2-11}
			& \multirow{3}{*}{LIDNN} & ListMLE & 0.427$^{+}$ & 0.219$^{+}$ & 0.427$^{+}$ & 0.301$^{+}$ & 0.435$^{+}$ & 0.325$^{+}$ & 0.455$^{+}$ & 0.344$^{+}$ \\ \cline{3-11}
			& & SoftRank & 0.457$^{+}$ & 0.234$^{+}$ & 0.442$^{+}$ & 0.314$^{+}$ & 0.445$^{+}$ & 0.336$^{+}$ & 0.464$^{+}$ & 0.355$^{+}$ \\ \cline{3-11}
			& & AttRank & 0.455$^{+}$ & 0.237$^{+}$ & 0.432$^{+}$ & 0.312$^{+}$ & 0.436$^{+}$ & 0.334$^{+}$ & 0.458$^{+}$ & 0.354$^{+}$ \\ \cline{2-11}
			& \multirow{3}{*}{\textbf{DLCM}} & ListMLE & 0.457$^{+}$ & 0.235$^{+}$ & 0.442$^{+}$ & 0.314$^{+}$ & 0.445$^{+}$ & 0.336$^{+}$ & 0.464$^{+}$ & 0.355$^{+}$ \\ \cline{3-11}
			& & SoftRank & \textbf{0.463}$^{*+\ddagger}$ & 0.243$^{*+\ddagger}$ & 0.444$^{*+\ddagger}$ & 0.320$^{*+\ddagger}$ & 0.447$^{*+\ddagger}$ & 0.342$^{*+\ddagger}$ & 0.465$^{*+\ddagger}$ & 0.360$^{*+\ddagger}$ \\ \cline{3-11}
			& & \textbf{AttRank} & \textbf{0.463}$^{*+\ddagger}$ & \textbf{0.246}$^{*+\ddagger}$ & \textbf{0.445}$^{*+\ddagger}$ & \textbf{0.322}$^{*+\ddagger}$ & \textbf{0.450}$^{*+\ddagger}$ & \textbf{0.344}$^{*+\ddagger}$ & \textbf{0.469}$^{*+\ddagger}$ & \textbf{0.362}$^{*+\ddagger}$ \\ \hline
			\hline
		\end{tabular}
	}
	\label{tab:results_MSLR30K}
\end{table*}

\begin{table*}[t]
	\centering
	\small
	\caption{Comparison of baselines and the DLCMs on Micrsoft 10K. $*$, $+$ and $\ddagger$ denotes significant improvements over the global ranking algorithm and the best corresponding re-ranking baseline (DNN) and LIDNN. 
	}
	\scalebox{1.0}{
			\begin{tabular}{ c | c | c || l | l | l | l | l | l | l | l    } 
			\hline
			\multicolumn{3}{c||}{ } & \multicolumn{8}{c}{Microsoft Letor Dataset 10K}\\ \hline 
			Initial List & Model & Loss Function & nDCG@1 & ERR@1 & nDCG@3 & ERR@3 & nDCG@5 & ERR@5 & nDCG@10 & ERR@10  \\\hline \hline
			\multicolumn{3}{c||}{SVMrank} & 0.292 & 0.129 & 0.312 & 0.199 & 0.329 & 0.226 & 0.360 & 0.248 \\ \hline \hline
			\multirow{9}{*}{SVMrank}&\multirow{3}{*}{DNN} & ListMLE  & 0.304$^{*\ddagger}$ & 0.134$^{*\ddagger}$ & 0.323$^{*\ddagger}$ & 0.208$^{*\ddagger}$ & 0.338$^{*\ddagger}$ & 0.234$^{*\ddagger}$ & 0.367$^{*\ddagger}$ & 0.256$^{*\ddagger}$ \\ \cline{3-11}
			& & SoftRank & 0.378$^{*\ddagger}$ & \textbf{0.207}$^{*\ddagger}$ & 0.366$^{*\ddagger}$ & 0.275$^{*\ddagger}$ & 0.368$^{*\ddagger}$ & 0.295$^{*\ddagger}$ & 0.386$^{*\ddagger}$ & 0.314$^{*\ddagger}$ \\ \cline{3-11}
			& & AttRank & 0.383$^{*\ddagger}$ & 0.203$^{*\ddagger}$ & 0.381$^{*\ddagger}$ & \textbf{0.276}$^{*\ddagger}$ & 0.388$^{*\ddagger}$ & \textbf{0.298}$^{*\ddagger}$ & 0.410$^{*\ddagger}$ & \textbf{0.318}$^{*\ddagger}$ \\ \cline{2-11}
			& \multirow{3}{*}{LIDNN} & ListMLE  & 0.283 & 0.125 & 0.305 & 0.197 & 0.320 & 0.222 & 0.355 & 0.245 \\ \cline{3-11}
			& & SoftRank & 0.295$^{*}$ & 0.130 & 0.311 & 0.201$^{*}$ & 0.328 & 0.227 & 0.358 & 0.249 \\ \cline{3-11}
			& & AttRank & 0.291 & 0.125 & 0.305 & 0.196 & 0.323 & 0.222 & 0.354 & 0.244 \\ \cline{2-11}
			& \multirow{3}{*}{\textbf{DLCM}} & ListMLE & 0.333$^{*\ddagger}$ & 0.152$^{*\ddagger}$ & 0.342$^{*\ddagger}$ & 0.225$^{*\ddagger}$ & 0.351$^{*\ddagger}$ & 0.249$^{*\ddagger}$ & 0.377$^{*\ddagger}$ & 0.271$^{*\ddagger}$ \\ \cline{3-11}
			& & \textbf{SoftRank} & \textbf{0.393}$^{*+\ddagger}$ & 0.205$^{*\ddagger}$ & \textbf{0.385}$^{*+\ddagger}$ & \textbf{0.276}$^{*\ddagger}$ & 0.388$^*\ddagger$ & \textbf{0.298}$^{*\ddagger}$ & 0.408$^{*\ddagger}$ & 0.317$^{*\ddagger}$ \\ \cline{3-11}
			& & \textbf{AttRank} &  0.390$^{*+\ddagger}$ & 0.206$^{*\ddagger}$ & 0.382$^{*\ddagger}$ & 0.275$^{*\ddagger}$ & \textbf{0.390}$^{*+\ddagger}$ & \textbf{0.298}$^{*\ddagger}$ & \textbf{0.411}$^{*\ddagger}$ & \textbf{0.318}$^{*\ddagger}$ \\ \hline \hline
			\multicolumn{3}{c||}{LambdaMART} & 0.419$^{+}$ & 0.223$^{+}$ & 0.417$^{+}$ & 0.302$^{+}$ & 0.425$^{+}$ & 0.325$^{+}$ & 0.446$^{+}$ & 0.344$^{+}$ \\ \hline \hline
			\multirow{9}{*}{LambdaMART} & \multirow{3}{*}{DNN} & ListMLE & 0.370 & 0.180 & 0.375 & 0.259 & 0.385 & 0.283 & 0.406 & 0.303 \\ \cline{3-11}
			& & SoftRank & 0.368 & 0.202 & 0.358 & 0.273 & 0.361 & 0.294 & 0.382 & 0.313 \\ \cline{3-11}
			& & AttRank & 0.378 & 0.198 & 0.373 & 0.272 & 0.380 & 0.295 & 0.403 & 0.314 \\ \cline{2-11}
			& \multirow{3}{*}{LIDNN} & ListMLE  & 0.414$^{+}$ & 0.220$^{+}$ & 0.413$^{+}$ & 0.299$^{+}$ & 0.422$^{+}$ & 0.323$^{+}$ & 0.441$^{+}$ & 0.342$^{+}$ \\ \cline{3-11}
			&& SoftRank & 0.420$^{+}$ & 0.223$^{+}$ & 0.415$^{+}$ & 0.301$^{+}$ & 0.425$^{+}$ & 0.325$^{+}$ & 0.445$^{+}$ & 0.344$^{+}$ \\ \cline{3-11}
			&& AttRank & 0.415$^{+}$ & 0.222$^{+}$ & 0.409$^{+}$ & 0.299$^{+}$ & 0.419$^{+}$ & 0.322$^{+}$ & 0.441$^{+}$ & 0.341$^{+}$ \\ \cline{2-11}
			& \multirow{3}{*}{\textbf{DLCM}} & ListMLE & 0.419$^{+}$ & 0.223$^{+}$ & 0.417$^{+\ddagger}$ & 0.302$^{+}$ & 0.425$^{+}$ & 0.325$^{+}$ & 0.446$^{+}$ & 0.344$^{+}$ \\ \cline{3-11}
			&& SoftRank & 0.425$^{*+\ddagger}$ & 0.230$^{*+\ddagger}$ & 0.419$^{*+\ddagger}$ & 0.306$^{*+\ddagger}$ & 0.426$^{+}$ & 0.329$^{*+\ddagger}$ & 0.447$^{+\ddagger}$ & 0.348$^{*+\ddagger}$ \\ \cline{3-11}
			&& \textbf{AttRank} & \textbf{0.432}$^{*+\ddagger}$ & \textbf{0.232}$^{*+\ddagger}$ & \textbf{0.423}$^{*+\ddagger}$ & \textbf{0.307}$^{*+\ddagger}$ & \textbf{0.429}$^{*+\ddagger}$ & \textbf{0.330}$^{*+\ddagger}$ & \textbf{0.450}$^{*+\ddagger}$ & \textbf{0.349}$^{*+\ddagger}$ \\ \hline
			\hline
		\end{tabular}
	}
	\label{tab:results_MSLR10K}
\end{table*}

\begin{table*}[t]
	\centering
	\small
	\caption{Comparison of baselines and the DLCMs on Yahoo! set 1. $*$, $+$ and $\ddagger$ denotes significant improvements over the global ranking algorithm and the best corresponding re-ranking baseline (DNN) and LIDNN. 
	}
	\scalebox{1.0}{
		\begin{tabular}{ c | c | c || l | l | l | l | l | l | l | l    } 
			\hline
			\multicolumn{3}{c||}{ } & \multicolumn{8}{c}{Yahoo! set 1}\\ \hline 
			Initial List & Model & Loss Function & nDCG@1 & ERR@1 & nDCG@3 & ERR@3 & nDCG@5 & ERR@5 & nDCG@10 & ERR@10  \\\hline \hline
			\multicolumn{3}{c||}{SVMrank} & 0.637 & 0.312 & 0.650 & 0.395 & 0.674 & 0.416 & 0.726 & 0.432 \\ \hline \hline
			\multirow{9}{*}{SVMrank}&\multirow{3}{*}{DNN} & ListMLE & 0.629 & 0.305 & 0.643 & 0.389 & 0.67 & 0.411 & 0.721 & 0.427 \\ \cline{3-11}
			&& SoftRank & 0.659$^{*\ddagger}$ & 0.337$^{*\ddagger}$ & 0.666$^{*\ddagger}$ & 0.412$^{*\ddagger}$ & 0.685$^{*\ddagger}$ & 0.432$^{*\ddagger}$ & 0.729$^{*\ddagger}$ & 0.447$^{*\ddagger}$ \\ \cline{3-11}
			&& AttRank & 0.667$^{*\ddagger}$ & 0.339$^{*\ddagger}$ & 0.675$^{*\ddagger}$ & 0.414$^{*\ddagger}$ & 0.695$^{*\ddagger}$ & 0.435$^{*\ddagger}$ & 0.740$^{*\ddagger}$ & 0.449$^{*\ddagger}$ \\ \cline{2-11}
			& \multirow{3}{*}{LIDNN} & ListMLE   & 0.540 & 0.247 & 0.611 & 0.357 & 0.641 & 0.379 & 0.698 & 0.395 \\ \cline{3-11}
			&& SoftRank & 0.642$^{*}$ & 0.319$^{*}$ & 0.647 & 0.398$^{*}$ & 0.670 & 0.419$^{*}$ & 0.717 & 0.435$^{*}$ \\ \cline{3-11}
			&& AttRank & 0.630 & 0.319$^{*}$ & 0.638 & 0.396$^{*}$ & 0.660 & 0.417$^{*}$ & 0.709 & 0.433$^{*}$ \\ \cline{2-11}
			& \multirow{3}{*}{\textbf{DLCM}} & ListMLE & 0.637 & 0.312 & 0.649 & 0.395 & 0.673 & 0.417$^{*}$ & 0.723 & 0.432 \\ \cline{3-11}
			&& SoftRank & 0.668$^{*+\ddagger}$ & 0.341$^{*+\ddagger}$ & 0.674$^{*\ddagger}$ & 0.416$^{*+\ddagger}$ & 0.694$^{*\ddagger}$ & 0.437$^{*+\ddagger}$ & 0.738$^{*\ddagger}$ & 0.451$^{*+\ddagger}$ \\ \cline{3-11}
			&& \textbf{AttRank} & \textbf{0.671}$^{*+\ddagger}$ & \textbf{0.342}$^{*+\ddagger}$ & \textbf{0.680}$^{*+\ddagger}$ & \textbf{0.417}$^{*+\ddagger}$ & \textbf{0.699}$^{*+\ddagger}$ & \textbf{0.438}$^{*+\ddagger}$ & \textbf{0.745}$^{*+\ddagger}$ & \textbf{0.452}$^{*+\ddagger}$ \\ \hline \hline
			\multicolumn{3}{c||}{LambdaMART} & 0.677$^{+}$ & 0.343$^{+}$ & 0.676$^{+}$ & 0.417$^{+}$ & 0.696$^{+}$ & 0.438$^{+}$ & 0.738$^{+}$ & 0.452$^{+}$ \\ \hline \hline
			\multirow{9}{*}{LambdaMART} & \multirow{3}{*}{DNN} & ListMLE & 0.640 & 0.314 & 0.653 & 0.398 & 0.678 & 0.419 & 0.727 & 0.434 \\ \cline{3-11}
			&& SoftRank & 0.641 & 0.330 & 0.643 & 0.404 & 0.663 & 0.425 & 0.711 & 0.440 \\ \cline{3-11}
			&& AttRank & 0.665 & 0.338 & 0.673 & 0.413 & 0.695 & 0.434 & 0.739 & 0.449 \\ \cline{2-11}
			& \multirow{3}{*}{LIDNN} & ListMLE & 0.603 & 0.298 & 0.621 & 0.384 & 0.651 & 0.407 & 0.71 & 0.423 \\ \cline{3-11}
			&& SoftRank & 0.677$^{+}$ & 0.343$^{+}$ & 0.676$^{+}$ & 0.417$^{+}$ & 0.694$^{+}$ & 0.438$^{+}$ & 0.738$^{+}$ & 0.452$^{+}$ \\ \cline{3-11}
			&& AttRank & 0.656 & 0.336 & 0.655 & 0.409 & 0.676 & 0.429 & 0.722 & 0.444 \\ \cline{2-11}
			& \multirow{3}{*}{\textbf{DLCM}} & ListMLE  & 0.667$^{+}$ & 0.334 & 0.673 & 0.412 & 0.692 & 0.433 & 0.736 & 0.447  \\ \cline{3-11}
			&& \textbf{SoftRank} & \textbf{0.678}$^{*+\ddagger}$ & \textbf{0.344}$^{*+\ddagger}$ & 0.678$^{*+\ddagger}$ & \textbf{0.418}$^{*+\ddagger}$ & 0.697$^{*+\ddagger}$ & \textbf{0.439}$^{*+\ddagger}$ & 0.739$^{*+}$ & \textbf{0.453}$^{*+\ddagger}$ \\ \cline{3-11}
			&& \textbf{AttRank} & 0.676$^{+}$ & 0.343$^{+}$ & \textbf{0.681}$^{*+\ddagger}$ & \textbf{0.418}$^{*+\ddagger}$ & \textbf{0.699}$^{*+\ddagger}$ & 0.438$^{+}$ & \textbf{0.743}$^{*+\ddagger}$ & \textbf{0.453}$^{*+\ddagger}$ \\ \hline
			\hline
		\end{tabular}
	}
	\label{tab:results_set1}
\end{table*}

\section{Results and Analysis}

In this section, we describe our results and conduct detailed analysis on the DLCM to show how it improves the ranking of existing learning-to-rank systems.

\subsection{Overall performance}


The overall retrieval performance of our baselines and corresponding DLCMs are shown in Table~\ref{tab:results_MSLR30K}, \ref{tab:results_MSLR10K} and \ref{tab:results_set1}. 
For each dataset, we split the baselines and our models into two groups.
Each group showed the results of one global ranking algorithm, three re-ranking baselines (ListMLE, SoftRank and AttRank), and the LIDNNs and DLCMs trained with different loss functions. 
As shown in Table~\ref{tab:results_MSLR30K}, re-ranking initial results using global ranking algorithms does not necessarily improve the performance of learning to rank systems.
When the initial ranker was weak (e.g. SVMrank), the re-ranking baselines (ListMLE, SoftRank and AttRank) produced better rankings for the initial results; when the initial ranker was strong (e.g. LambdaMART), however, the re-ranking baselines actually hurt the performance of the whole system.
Although the re-ranking baselines in our experiments adapted listwise loss functions, they share the same global assumption with the global ranking algorithms used in the initial retrieval but were trained only with the top results, which is a limited subset of the training corpora.
Therefore, the re-ranking baselines do not incorporate any new information and could harm the ranking systems. 

According to our experiments, directly applying deep models on the concatenation of all document features did not work well.
None of the LIDNN models consistently outperformed their corresponding initial rankers.
In fact, the results of the LIDNNs were highly correlated with the performance of their initial rankers. 
In Table~\ref{tab:results_MSLR30K}, the LIDNN with SVMrank performed even worse than the re-ranking baselines.
The number of parameters in LIDNN is usually large (more than 1 million) due to the large amount of input features, but this doesn't make it powerful empirically.
One possible explanation is that concatenating all document features together makes it difficult to discriminate the relevance of individual documents in fine granularity. 
As a result, the LIDNNs just learned to fit the initial ranking of the inputs.

In contrast to the baseline models, re-ranking with the DLCMs brought stable and significant improvements to all of the global ranking algorithms. 
On SVMrank in Microsoft 30K, the DLCM with SoftRank achieved 40.9\%, 80.6\% improvements on NDCG@1, ERR@1 and 15.9\%, 35.8\% improvements on NDCG@10, ERR@10.
On LambdaMART, which is considered to be one of the state-of-the-art learning-to-rank models, the DLCM with AttRank achieved 1.3\%, 4.7\% improvements on NDCG@1, ERR@1 and 1.1\%, 2.0\% improvements on NDCG@10, ERR@10.
Although the number of trainable parameters in DLCMs is close to the re-ranking baselines and much lower than LIDNNs, the DCLMs significantly outperformed them in our experiments.
This indicates that incorporating local ranking context with the DLCM is beneficial for global ranking algorithms.

For different variations, the DLCMs with AttRank consistently produced better results than the DLCMs with ListMLE and outperformed its SoftRank version on Microsoft 30K LambdaMART, Microsoft 10K LambdaMART and Yahoo! SVMrank.
Because the Attention Rank loss is much simpler and more efficient than the ListMLE (2 times faster) and the SoftRank (20 times faster) empirically, we believe that it has great potentials in real applications.

Compared to other datasets, we notice that the improvements from the DLCMs are relatively small on Yahoo! Letor set 1.
This, however, is not surprising considering the special properties of the Yahoo! data.
First, Yahoo! Letor set 1 is a relatively easy dataset and the ranked lists produced by the baseline methods are nearly perfect (e.g. LambdaMART had 0.738 on NDCG@10).
When we input those nearly perfect ranked lists into the DLCM, it is less likely to learn anything new other than producing the same ranking from the initial ranked list.  
Second, every feature in Yahoo! Letor set 1 is a high-quality ranking signal in itself. 
The 700 features in Yahoo! data are the outputs of a feature selection where the most predictive features for ranking are kept~\cite{chapelle2011yahoo}. 
This means that each feature already has a high correlation with the document's relevance label in global.
The information from the local ranking context is unlikely to be useful on Yahoo! letor set 1 given the fact that all features are selected globally. 



\subsection{Pair-wise Ranking Analysis}


\begin{table}
	\centering
	\small
	\caption{The statistics of the test fold used for pairwise ranking analysis in Microsoft 30k. \textit{Query} denotes the number of queries containing documents with the corresponding label.}
	\begin{tabular}{ l | l l l l l | l    } 
		\hline
		Rel. Label & 0 & 1 & 2 & 3 & 4 & All\\ \hline
		Doc. & 104k & 79k & 47k & 8k & 4k & 241k \\
		Query & 6,285 & 6,069 & 5,605 & 3,209 & 1,612 & 6,306 \\
		\hline
	\end{tabular}
	\label{tab:fold_data}
\end{table}

In this section, we want to shed some light on how our DLCMs improve the global ranking baselines.
Our analysis focused on two questions:
(1) Compared to baselines, do all relevant documents receive rank promotions in our DLCMs?
(2) What queries received more improvements from the use of local ranking context? 


For analysis purposes, we used the ranked lists of LambdaMART and its DLCM with AttRank from one test fold of Microsoft 30k and analyzed the pairwise ranking changes on documents with different relevance labels $y_{(q,d)}$.
The statistics of the test fold are shown in Table~\ref{tab:fold_data}.
We measured the improvement of the DLCM with AttRank over LambdaMART on a document $d$ by counting the reduction of negative ranking pairs.
A negative ranking pair (\textit{NegPair}) is a document pair $(d, d')$ where $d'$ is ranked higher than $d$ in query $q$ while $y_{(q,d')} < y_{(q,d)}$. 
In other words, the number of negative pairs is the number of documents which have been incorrectly placed in front of $d$. 
For simplicity, we use $NP(d, LambdaMART)$ and $NP(d,DLCM)$ to denote the negative ranking pair of $d$ in the ranked lists of LambdaMART and DLCM with AttRank.
The \textit{NegPair reduction} is defined as $NP(d, LambdaMART)-NP(d,DLCM)$.


\begin{figure}
	\centering
	\includegraphics[width=2.65in]{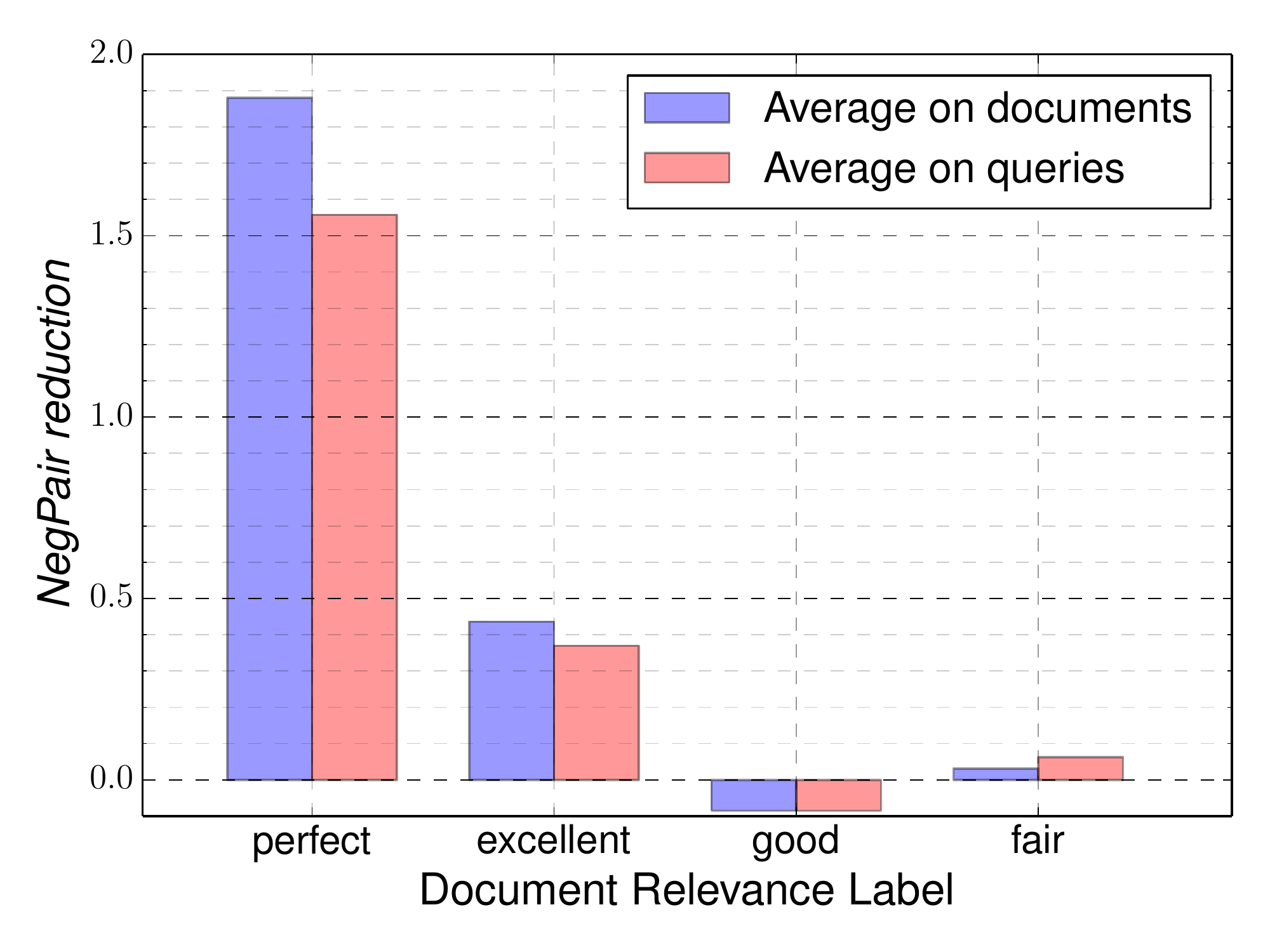}
	\caption{The \textit{NegPair reduction} ($NP(d, LambdaMART)$-$NP(d,DLCM)$) on documents with different relevance labels.}
	\vspace{-5pt}
	\label{fig:doc_pair}
\end{figure}

\begin{figure}
	\centering
	\includegraphics[width=2.65in]{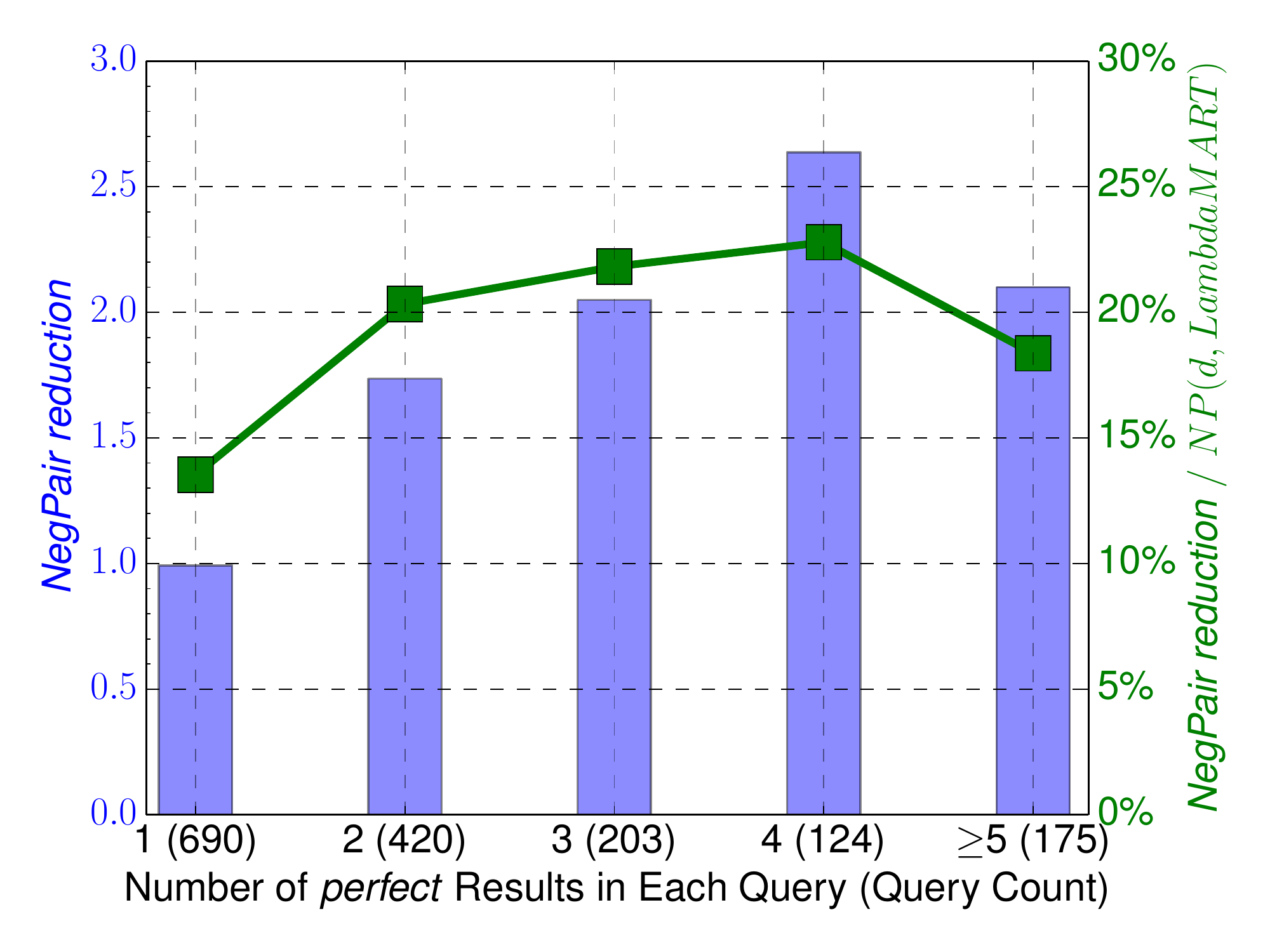}
	\caption{The \textit{NegPair reduction} and corresponding improvement proportion for queries with different number of \textit{perfect} documents.}
	\label{fig:query_pair}
\end{figure}

\begin{figure*}
	\centering
	\begin{subfigure}{.25\textwidth}
		\centering
		\includegraphics[width=1.81in]{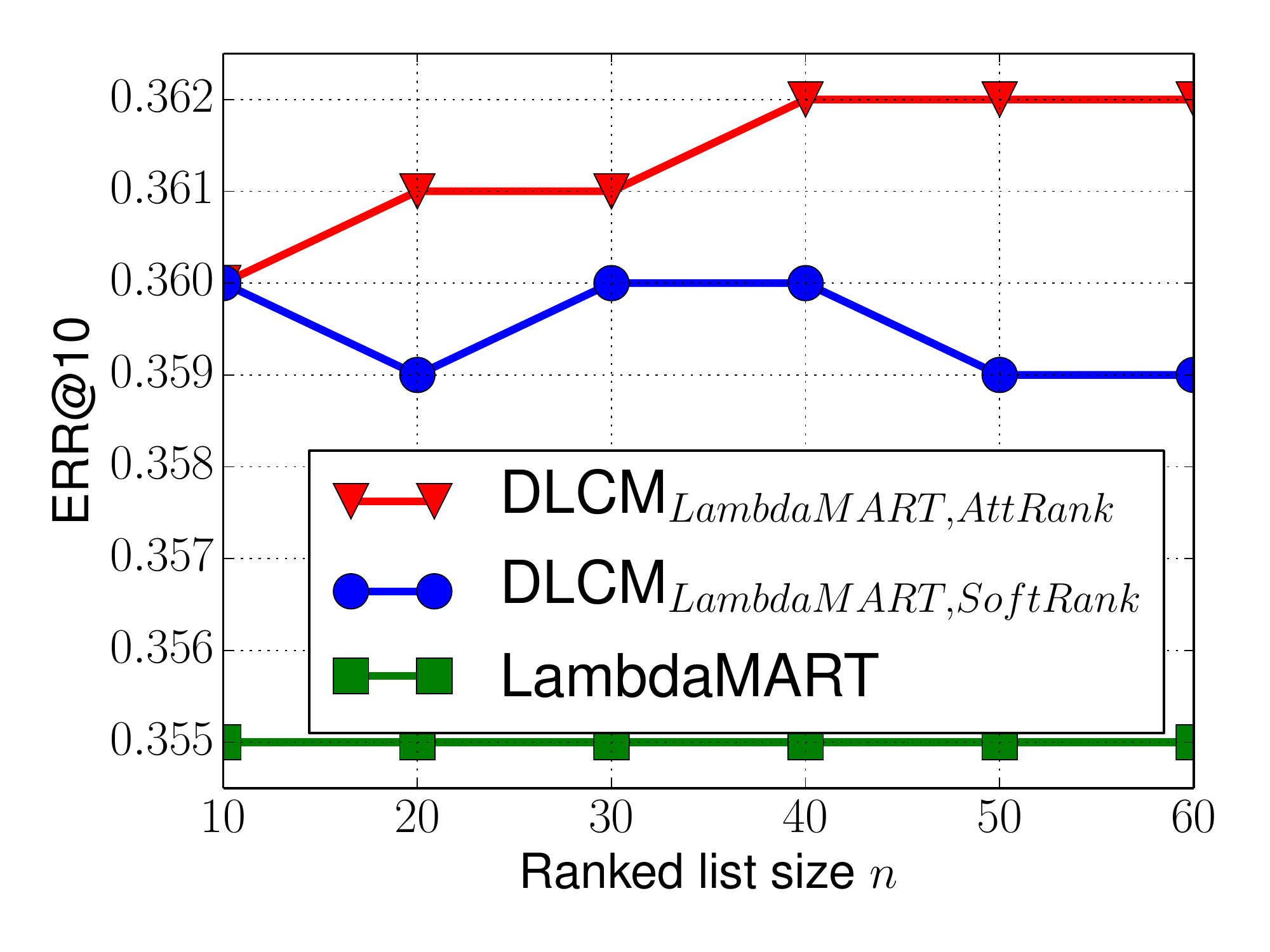}
		\caption{Input ranked list size $n$}
		\label{fig:rerank_size}
	\end{subfigure}%
	\begin{subfigure}{.25\textwidth}
		\centering
		\includegraphics[width=1.81in]{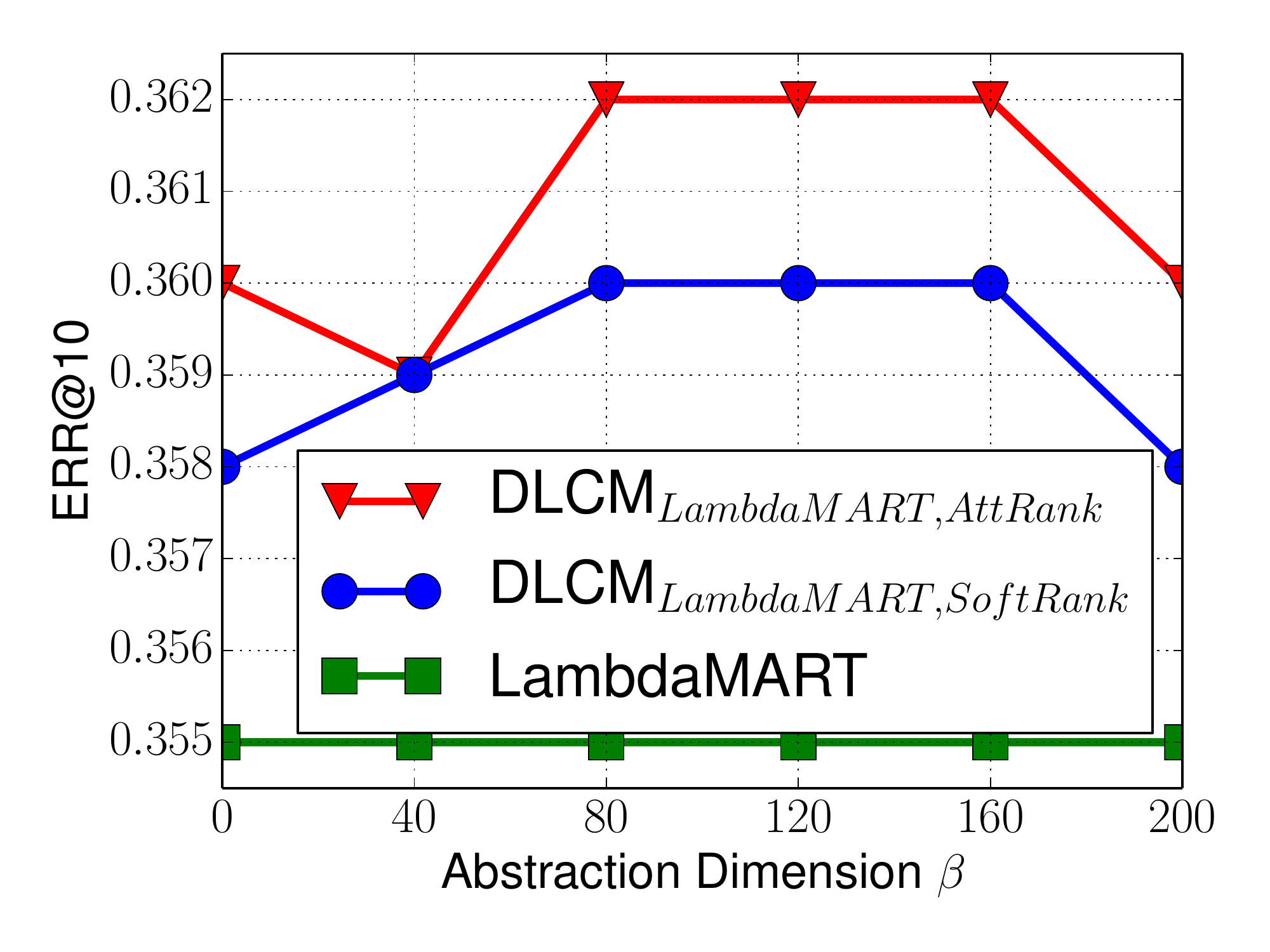}
		\caption{Abstraction dimension $\beta$}
		\label{fig:beta}
	\end{subfigure}%
	\begin{subfigure}{.25\textwidth}
		\centering
		\includegraphics[width=1.81in]{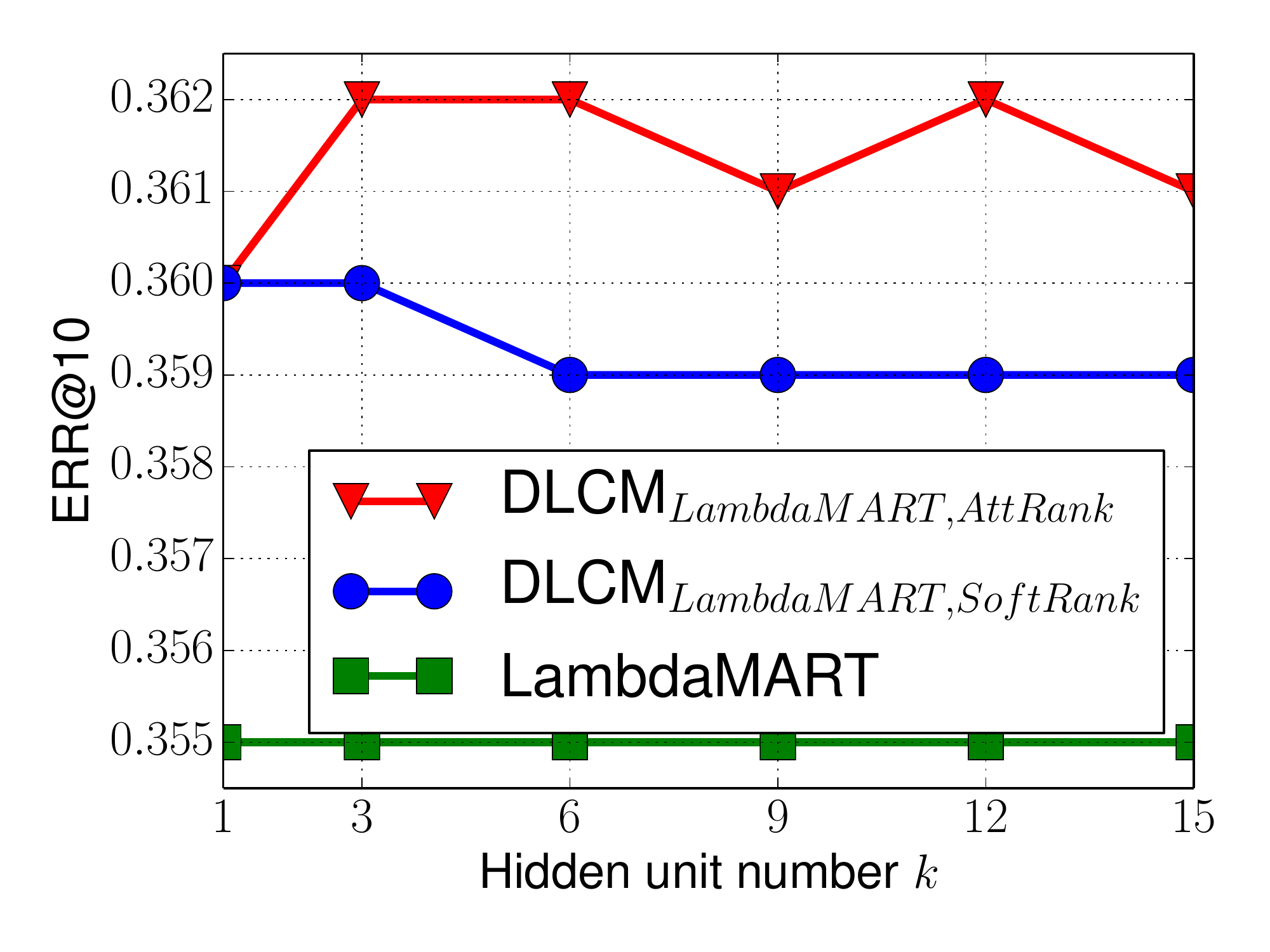}
		\caption{Hidden unit number $k$}
		\label{fig:num_heads}
	\end{subfigure}%
	\begin{subfigure}{.25\textwidth}
		\centering
		\includegraphics[width=1.81in]{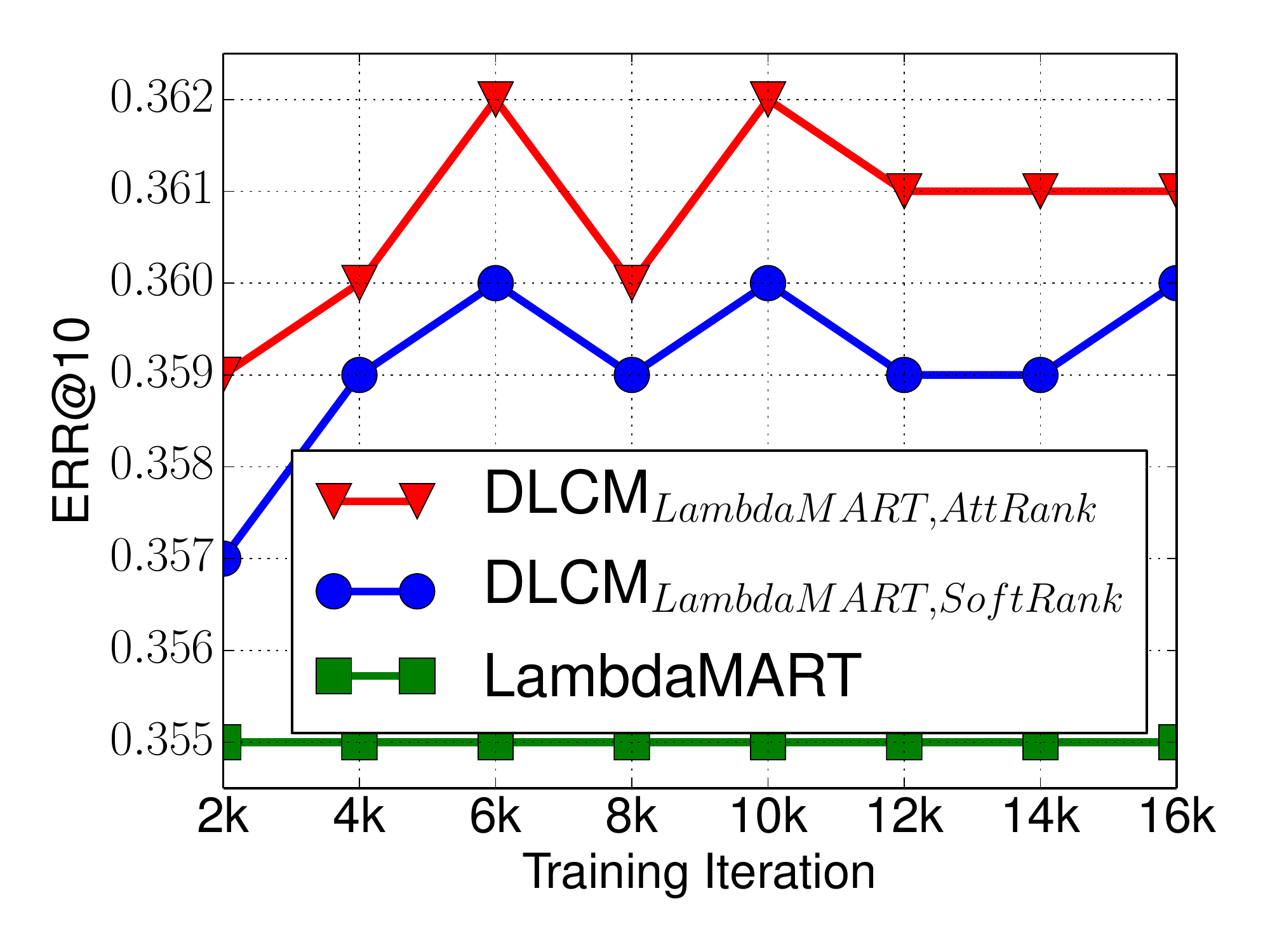}
		\caption{Training iteration}
		\label{fig:iteration}
	\end{subfigure}%
	\caption{The performance of the DLCMs on Microsoft 30k with different hyper-parameters.}
	\label{fig:parameter}
\end{figure*}

To answer the first question, we plotted the average \textit{NegPair reduction} for documents with different relevance labels in Figure~\ref{fig:doc_pair}.
The numbers are averaged both on document level and on query level (average on documents within each query and then on all the queries).
As shown in Figure~\ref{fig:doc_pair}, the  \textit{perfect} results (documents with label 4) received the largest promotions in rank.
The average \textit{NegPair reduction} per document on \textit{perfect} is 1.88, which means that the positions of these documents have been effectively increased by nearly 2 in their ranked lists.
The \textit{NegPair reduction} on \textit{excellent} results (label 3) are smaller but still promising (0.43 per document). 
On \textit{good} and \textit{fair} results (label 2 and label 1), however, there is almost no improvement when comparing the DLCM with LambdaMART. 
These observations indicate that our DLCMs are particularly good at finding \textit{perfect} and \textit{excellent} results from the initial ranked list, but not capable of discriminating fair results from irrelevant ones.

To answer the second question, we showed the distribution of \textit{NegPair reduction} for \textit{perfect} documents by splitting queries according to how many \textit{perfect} results they have.
In Figure~\ref{fig:query_pair}, the horizontal axis represents the number of \textit{perfect} results in each query (and the count of queries within that category) and the vertical axises represent both the average \textit{NegPair reduction} ($NP(d, LambdaMART)-NP(d,DLCM)$) and corresponding improvement proportion (e.g. \textit{NegPair reduction} is 2.64 and $NP(d,LambdaMART)$ is 11.57 on queries with four \textit{perfect} documents, so the improvement proportion is 22.79\%).
The overall results in Figure~\ref{fig:query_pair} show that the \textit{NegPair reduction} gradually increases when the query has more \textit{perfect} results.
The \textit{NegPair reduction} is 0.99 (13.5\%) on queries with one \textit{perfect} document but 2.64 (22.8\%) on queries with four \textit{perfect} documents.
These numbers imply that the DLCMs are more useful in queries with more than one \textit{perfect} result in the original ranked list.
This basically supports the hypothesis that relevance information from top results can help us better predict the relevance of each document with learning-to-rank models.

\subsection{Parameter Sensitivity}\label{sec:parameter_analysis}

To evaluate the parameter sensitivity of DLCM, we tested the DLCM with SoftRank (DLCM$_{LambdaMART, SoftRank}$) and AttRank (DLCM$_{LambdaMART, AttRank}$) with different hyper-parameters on Microsoft 30K. 

Figure~\ref{fig:rerank_size} shows the ERR@10 of the DLCMs with different input ranked list size $n$.
As shown in Figure~\ref{fig:rerank_size}, the performance of the DLCM with AttRank became better when $n$ increased from 10 to 40 and then remained unchanged when $n$ increased from 40 to 60.
In comparison, the performance of the DLCM with SoftRank showed no clear correlation with the size of input ranked lists.
In the beginning, when $n$ is relatively small, the increase of $n$ would introduce more relevant documents into the inputs of DLCMs and give the model more opportunities to improve the ranked list.
When $n$ is large enough, however, increasing $n$ only brings irrelevant documents into the initial ranked list. 
Compared to the DLCM with SoftRank, the DLCM with AttRank is more robust to noise and more capable of gathering relevant information from the initial ranked lists.

Figure~\ref{fig:beta} depicts the results of the DLCM with different abstraction dimension $\beta$.
When $\beta$ equals to 0, it means that there is no abstraction process and we used the original feature vector from the training data as our model inputs.
Overall, adding abstraction layers was beneficial for the performance of DLCMs.
The abstraction process introduces more parameters into the DLCM and has the potential to improve the model robustness with respect to input noise.
Nonetheless, large $\beta$ does not necessarily lead to better performance.
The performance of the DLCM with $\beta=200$ has no significant difference with the performance of the DLCM with $\beta=0$ in our experiments.
Because the original feature vector has 136 dimensions, adding an abstraction with larger dimensions than the original input may not be reasonable and could bring unnecessary computation cost to the training of DLCMs.

In Figure~\ref{fig:num_heads} and Figure~\ref{fig:iteration}, the performance of the DLCM is relatively stable after $k$ and training iterations are larger than certain thresholds.
The improvement on ERR@10 is not noticeable when we increased $k$ from 3 to 15 and the performance of the DLCM fluctuated before 12,000 iterations, but remained stable afterward.




\section{Conclusion and Future Work}

In this paper, we propose a Deep Listwise Context Model to improve learning-to-rank systems with the local ranking context.
Our model uses a RNN to encode the top retrieved documents from a global learning-to-rank algorithm and refines the ranked list with a local context model.
Our model can be efficiently trained with our attention-based listwise ranking loss and directly deployed over existing learning-to-rank models without additional feature extraction or retrieval processing.
In our experiments, we showed that our DLCM can significantly improve the performance of baseline methods on benchmark learning-to-rank datasets.
Also, our analysis indicates that the DLCM is particularly good at finding the best documents from the initial ranked list.
These results support the hypothesis that the local ranking context from top retrieved documents are valuable for learning to rank, which potentially provides new ideas for the future study.

As the DLCM learns to rank documents according to the local ranking context of top results, there is a concern as to whether the model would place similar documents in high positions and hurt ranking diversity. 
We did not discuss this problem in this paper, but it could be another fruitful research direction in the future.
We believe that our framework has the potential to improve search result diversity as well.
For example, we could add a decoding phase into our model and generate each result according to both the ranking features and previous outputs.
This process resembles the paradigm used by many ranking diversity models.
We will further explore this direction in our future work.

\iftrue
\section{Acknowledgments}
This work was supported in part by the Center for Intelligent Information Retrieval and in part by NSF IIS-1160894. Any opinions, findings and conclusions or recommendations expressed in this material are those of the authors and do not necessarily reflect those of the sponsor.
\fi

\balance
\bibliographystyle{ACM-Reference-Format}
\bibliography{sigproc} 

\end{document}